\documentclass[twocolumn,trackchanges]{aastex61}

\received{October 09, 2016}
\revised{February 07, 2017}
\accepted{February 08, 2017}
\submitjournal{ApJ}

\shorttitle{location of $\gamma$-ray emission region of GB 1310+487}
\shortauthors{Shi-Ju Kang}
\begin{document}

%\title{Constraining the location of gamma-ray emission region of the gamma-ray-loud radio source GB 1310+487}
\title{Constraints on the location of the gamma-ray emission region for the gamma-ray-loud radio source GB 1310+487}

\correspondingauthor{Shi-Ju Kang}
\email{kangshiju@hust.edu.cn}

\author[0000-0002-9071-5469]{Shi-Ju Kang}
\affil{Department of Physics and Electronics Science, Liupanshui Normal University, Liupanshui, Guizhou, 553004, China}

\begin{abstract}

We employ a single-zone leptonic jet model, with synchrotron, synchrotron self-Compton (SSC) and external Compton (EC) process, to reproduce the quasi-simultaneous multi-wavelength spectral energy distributions in active and quiescent states of the narrow-line gamma-ray-loud radio source GB 1310+487. {In the case of the EC process}, the external seed photons from both broad line region (BLR) and dust torus are considered by assuming that the gamma-ray emission {region is located at} the outside boundary of the BLR and inside the dust torus. Comparing the energy density of external {photon} fields $U_{\rm BLR}$ obtained by model fitting with that constrained from the BLR observations. We find that the location of the gamma-ray emitting region of GB 1310+487 can be tightly constrained at the outer edge of the BLR (the dissipation distance of the $\gamma$-ray emission region from central black hole  $r_{\rm diss} \sim {\rm~a~few~times~of}~R_{\rm BLR}$). The ratio of magnetic energy and emitting-electron energy in the radiation blob ($\epsilon_{B}=L_{B}/L_{\rm e}$) is gradually increased  from Flare 1, Flare 2 to Post-Flare, where the magnetic energy increase and matter energy decrease. These results suggest that the conversion of the magnetic field and the matter (radiation electrons) energy and the location of the $\gamma$-ray emission region (or ambient photon field) may play an important role in different radiation states of GB 1310+487.

\end{abstract}





\section{Introduction} \label{sec:intro}

Blazars, including flat-spectrum radio quasars (FSRQs) and BL Lacertae objects (BL Lacs), are a peculiar sub-class of radio-loud active galactic nuclei ({AGN}) with a relativistic jet pointed at a small viewing angle to the line of sight \citep{1995PASP..107..803U}. The multi-wavelength spectral energy distributions (SEDs) from radio to $\gamma$-ray bands of blazars dominantly come from non-thermal emission, where the SED normally exhibits a two-hump structure in the $\nu-\nu F_{\nu}$ space. The lower energy hump (peaked at between mm and soft X-ray wavelengths) is normally attributed to the synchrotron emission produced by the non-thermal electrons in the jet while the second hump (peak at the MeV-GeV range) mainly come from inverse Compton (IC) scattering. The seed photons for IC scattering may come from the synchrotron {photons inside the jet} (SSC process, e.g., \citealt{1981ApJ...243..700K}; \citealt{1985ApJ...298..114M}; \citealt{1989ApJ...340..181G}) and/or external photons (EC process) {from outside the jet}, where the external photons possibly originate from the accretion disk (e.g., \citealt{1993ApJ...416..458D}; \citealt{1997A&A...324..395B}), the broad line region (BLR; e.g., \citealt{1994ApJ...421..153S}; \citealt{1996MNRAS.280...67G}), and/or the molecular torus (e.g., \citealt{2000ApJ...545..107B}; \citealt{2008MNRAS.387.1669G}). {A pure SSC model} was widely adopted in fitting the multi-wavelength SED of  high-synchrotron-peaked (HSP, \citealt{2010ApJ...716...30A}) BL Lacs (e.g., \citealt{1997A&A...320...19M}; \citealt{2004ApJ...601..151K}; \citealt{2014ApJ...788..104Z}), while luminous FSRQs prefer SSC+EC model (e.g., \citealt{1999ApJ...515..140S}; \citealt{2002ApJ...581..127B}; \citealt{2011ApJ...735..108C}; \citealt{2014MNRAS.439.2933Y}).

%%%%%%%%%%%%%%%%%%%%%%%%%%%%%%%%%%%%%%%%%%%%%%%%%%%%%%%%%%%%%%

  GB 1310+487 is an extragalactic flat-spectrum radio source with a redshift $z = 0.638$ in the Fermi $\gamma$-ray source catalogs, listed as 1FGL J1312.4+4827, 2FGL J1312.8+4828 and 3FGL J1312.7+4828 in the First, Second and Third Fermi-LAT catalog (1FGL; \citealt{2010ApJS..188..405A}, 2FGL; \citealt{2012ApJS..199...31N} and 3FGL; \citealt{{2015ApJS..218...23A}}) respectively. A gamma-ray flare was observed by the Fermi Large Area Telescope on 2009 November 18 (LAT; \citealt{2009ApJ...697.1071A}), with a daily flux of $\sim 10^{-6}$ photons ${\rm cm}^{-2} {\rm s}^{-1}$ at energies E $>$ 100 MeV (\citealt{2009ATel.2306....1S,2014A&A...565A..26S}), then it became one of the brightest GeV $\gamma$-ray sources for about two weeks.

The multi-wavelength SEDs of GB 1310+487 show a double-peaked structure \citep{2014A&A...565A..26S}, which is the typical features of blazars and gamma-ray-loud narrow-line Seyfert 1 (NLSy1) galaxies. The multi-wavelength SEDs of the three different states of GB 1310+487 were organized, and the evolution of the observed SEDs were preliminarily discussed by \cite{2014A&A...565A..26S} {on the basis of a} blazar leptonic jet model, which has successfully been used to explain the SEDs of blazar, and also was employed to study the SED of GeV-bright NLSy1 galaxies (e.g., \citealt{2009ApJ...707L.142A}; \citealt{2013ApJ...768...52P}; \citealt{2015ApJ...798...43S}; \citealt{2016ApJ...819..121P}; \citealt{2016ApJ...820...52P}) and non-blazar GeV-bright AGNs, such as Perseus A (NGC1275; \citealt{2009ApJ...699...31A}), M87 (\citealt{2009ApJ...707...55A}), Cen A (\citealt{2001MNRAS.324L..33C}) and 3C 120 (e.g., \citealt{2015A&A...574A..88S}).
\cite{2014A&A...565A..26S} proposed that the GeV $\gamma$-ray emission of GB 1310+487 is dominated by the EC process. 
However, the source of the seed photons for the EC process is not determined due to the unclear  $\gamma$-ray emission region location, where the external seed photons may come from accretion-disk {and/or  BLR, and/or} dusty torus. Furthermore, the external photons may also come from multiple components. The $\gamma$-ray spectrum is a varying contribution from multiple EC components (e.g., EC on accretion disk and dusty torus photons or EC on BLR and dusty torus photons{; e.g.,} \citealt{2013MNRAS.431..824B}; \citealt{2010ApJ...714L.303F}; \citealt{2014ApJ...782...82D}; \citealt{2015MNRAS.454.1310Y}; \citealt{2015ApJ...803...15P}; \citealt{2017ApJS..228....1Z}). \cite{2016PASP..128d4101Y} {employed a synchrotron} + SSC + EC model to investigate the gamma-ray origin of the GeV-bright active galaxy GB 1310+487 through modeling its quasi-simultaneous SEDs in active and quiescent states. They proposed that the GeV gamma-ray emission of GB 1310+487 is {dominated by the EC process} scattering external soft photons {coming from} a simple blackbody radiation spectrum with a characteristic temperature $T_{\rm ext}\sim11.2$ eV.

Some recent works suggested that the high energy gamma rays might {come from multiple emission regions} (e.g., \citealt{2013MNRAS.431..824B}) or external soft photons come from two/multiple emission region (e.g., BLR and dusty torus) in {an} EC process (e.g., \citealt{2010ApJ...714L.303F}; \citealt{2014ApJ...782...82D}; \citealt{2015MNRAS.454.1310Y}; \citealt{2015ApJ...803...15P}). In order to understand the possible origin of the gamma rays in GB 1310+487, in this work, we try to explore whether the traditional one-zone leptonic model after including a multiple EC components, where external soft photons come from both BLR and dusty torus, can explain its multi-wavelength SEDs or not. Throughout the letter, we assume the following cosmology: $H_{0}=70\ \rm km\ s^{-1} Mpc^{-1}$, $\Omega_{0}=0.3$ and $\Omega_{\Lambda}=0.7$.

%%%%%%%%%%%%%%%%%%%%%%%%%%%%%%%%%%%%%%%%%%%%%%%%%%%%%%%%%%%%%%

\section{The Model}\label{sec:model}

In this work, we adopt the traditional one-zone synchrotron + IC model to fit the SEDs of GB 1310+487, a model that is widely used in blazars (see e.g., \citealt{2010MNRAS.402..497G} and references therein). A homogeneous sphere with radius $R$ embedded in a magnetic field $B$ is assumed, that moves relativistically with a speed of $\upsilon=\beta c$ ($c$ is the speed of light in vacuum, {bulk Lorentz factor} $\Gamma=1/\sqrt{1-\beta^2}$) along the jet orientation. Doppler factor $\delta=\left[\Gamma\left(1-\beta\cos\theta\right)\right]^{-1}\approx\Gamma$ is assumed for the relativistic jet with a small viewing angle $\theta\leq 1/\Gamma$. The electron spectrum is assumed as a broken power-law distribution, with {indices} $p_{1}$ and $p_{2}$ below and above the break energy $\gamma_{b}m_{e}c^{2}$,
   \begin{equation}
   N(\gamma )=\left\{ \begin{array}{ll}
                    N_{0}\gamma ^{-p_1}  &  \mbox{ $\gamma_{\rm min}\leq \gamma \leq \gamma_{\rm b}$} \\
            N_{0}\gamma _{\rm b}^{p_2-p_1} \gamma ^{-p_2}  &  \mbox{ $\gamma _{\rm b}<\gamma\leq\gamma_{\rm max}$}
           \end{array}
       \right.
  \label{Ngamma}
  \end{equation}
where $\gamma_{\rm min}$ and $\gamma_{\rm max}$ are the minimum and maximum electron Lorentz factors, and $N_{0}$ is the normalization of the particle distribution. Such a broken power-law distribution is a steady-state electron spectrum, which could be the result of the balance between the particle cooling and escape rates in the blob (e.g., \citealt{1962SvA.....6..317K}; \citealt{1994ApJ...421..153S}; \citealt{1996ApJ...463..555I}; \citealt{1998A&A...333..452K}; \citealt{1998MNRAS.301..451G}; \citealt{2002ApJ...581..127B}; \citealt{2012ApJ...748..119C}; \citealt{2013ApJ...768...54B}).

  Some recent works suggested that the $\gamma$-ray emission region of {blazar jets} might be located near the outer boundary of the BLR and within the dust tours (e.g., {\citealt{2012A&A...537A..70S,2013ApJ...773..147J}; \citealt{2013ApJ...771L...4C}; \citealt{2014ApJ...782...82D}; \citealt{2015ApJ...808..162C}; \citealt{2016ApJ...821..102B};   \citealt{2017ApJS..228....1Z}), where contributions from both BLR and torus photons are required to explain the observed gamma-ray spectrum. {In the EC process the external soft photons come} from two emission region (e.g., both BLR and dusty torus; e.g., \citealt{2010ApJ...714L.303F}; \citealt{2014ApJ...782...82D}; \citealt{2015MNRAS.454.1310Y}; \citealt{2015ApJ...803...15P}).
  Since the location of the $\gamma$-ray emission region is still unclear, {different from \cite{2016PASP..128d4101Y}  the external seed photons are considered to originate} from one single region (e.g., BLR or dust tours), we assume a dual-component Compton-scattering scenario {in which the external} seed photons predominantly originate from both the BLR and the dust torus, where the gamma-ray emission region locate outside the broad-line region and within the dusty torus.

  The external radiation field is characterized by an isotropic blackbody with the temperature $T=h\nu_{\rm p}/(3.93k_B)$, where $\nu_{\rm p}$ is the peak frequency of seed photons in the $\nu-\nu F_{\nu}$ space. For the BLR cloud, the most prominent contribution comes from the Ly$\alpha$ line, and hence the spectrum is assumed to be a blackbody with a peak around $2\times10^{15}~\Gamma$ Hz \citep[see,][]{2008MNRAS.387.1669G}. For the IR torus, the spectrum is assumed to be a blackbody with a peak frequency of $\nu_{\rm IR} = 3 \times 10^{13}~\Gamma$ Hz in the comoving frame \citep{2007ApJ...660..117C}.
  The energy densities of external photon {fields of the BLR ($U_{\rm BLR}$) and the dusty} torus ($U_{\rm torus}$) are a function of the distance from the central black hole (e.g., \citealt{2009MNRAS.397..985G}; \citealt{2009ApJ...704...38S}; \citealt{2012ApJ...754..114H}). Assuming the gamma-ray emission region {is located outside the BLR} and within the dusty torus, the value of the $U_{\rm BLR}$ decreases quickly, while the the value of the $U_{\rm torus}$ is roughly {not changed}. So in the model, the $U_{\rm BLR}$ is set as a free parameter and the $U_{\rm torus}=3\times10^{-4}\Gamma^{2}$ erg cm$^{-3}$ \citep{2007ApJ...660..117C} is assumed in the jet comoving frame.

  The Klein-Nishina effect in the inverse Compton scattering and the self-absorption effect in synchrotron emission are properly considered \citep[see,][]{1979rpa..book.....R, 1970RvMP...42..237B}. The high energy $\gamma$-ray emission is expected to be significantly absorbed by the extragalactic background light (EBL) via pair production. The absorption of gamma-rays by the EBL can be estimated using the model-dependent gamma-ray opacity of $\tau(\nu,z)$, where the relation between the observed spectrum, ${F}_{\rm obs}(\nu)$, and the intrinsic spectrum, ${F}_{\rm in}(\nu)$, can be described as follows:
  \begin{equation}
  F_{\rm obs}(\nu,z) = e^{-\tau(\nu,z)}F_{\rm in}(\nu,z),
  \end{equation}
  where $\tau(\nu,z)$ is the absorption optical depth resulting from interactions with the EBL (e.g., \citealt{{2004A&A...413..807K}}; \citealt{2005ApJ...618..657D};
   \citealt{2012MNRAS.422.3189G}; \citealt{2008A&A...487..837F}; \citealt{2010ApJ...712..238F}; \citealt{2010A&A...515A..19K}; \citealt{2011MNRAS.410.2556D}; \citealt{2012Sci...338.1190A}; \citealt{2013A&A...550A...4H}). In order to minimize hardening introduced from EBL absorption corrections, we adopt the absorption optical depth derived from the EBL model proposed by \cite{2011MNRAS.410.2556D} in our calculations. In our SED modeling of {Figure~\ref{fig:flare}}, we assume the model prediction as the intrinsic emission and correct it to our local universe using equation (2), and compare it with the observational data (e.g., \citealt{2011ApJ...728..105Z}; \citealt{2013ApJ...764..113Z}; \citealt{2013MNRAS.431.2356Z,2014MNRAS.442.3166Z,2016A&A...585A...8Z}; \citealt{2012ChA&A..36..115K,2014JApA...35..385K,2014ApJS..215....5K,2016MNRAS.461.1862K}).

\section{Model the SEDs of GB 1310+487}\label{sec:fitting}

  The quasi-simultaneous multi-wavelength data from high energy {gamma-rays} ($Fermi$-LAT, AGILE), X-ray and UV ($Swift$), optical (Kanata, NOT, and Keck), infrared (IR, OAGH and WISE) and radio (IRAM 30m, OVRO 40m, Effelsberg 100 m, RATAN-600 and VLBA) for GB 1310+487 at two active states (Flare 1 and Flare 2) and Post-Flare state are collected from \cite{2014A&A...565A..26S} and shown in Figure~\ref{fig:flare}. Flare 1 (the first and brighter flare) showed a higher flux and peaked around 2009 November 27 ({JD2455163}) with the weekly averaged flux of $(1.4 \pm 0.1)\times10^{-6}$ photons $\rm cm^{-2}~s^{-1}$. Flare 2 (the second flare ) showed a weekly integrated flux of $(0.54 \pm 0.07)\times10^{-6}$ photons $\rm cm^{-2}~s^{-1}$  and peaked around 2010 June 17 ({JD2455365}), with the daily flux of $\sim 0.54 \times10^{-6}$ photons $\rm cm^{-2}~s^{-1}$ lasting about three weeks. {The two flares present various kinds of flux evolution.} Flare 1 shows a fast rise and slower decay trending, while a gradual flux rise and rapid decay was observed in Flare 2 (see, \citealt{2014A&A...565A..26S}).

  We apply the one-zone jet model as described in Section 2 to reproduce the multi-waveband SEDs of GB 1310+487. There are 10 parameters in the SSC + EC (BLR) + EC (torus) model: $R$, $\delta$, $B$, $p_{1}$, $p_{2}$, $\gamma_{\rm min}$, $\gamma_{\rm max}$, $\gamma_{\rm b}$, $N_{0}$ {and} $U_{\rm BLR}$. In order to reduce the number of free parameters, the radius of the emitting {region in the jet} frame can be {constrained with the} minimum variability timescale and redshift with $R \leqslant\delta c t_{\rm var}/(1+z)\sim 1.58 \times 10^{15} \delta$ cm, where the timescale of variability of the $\gamma$-ray observations of Fermi is about half a day \citep{2014A&A...565A..26S}. {A conservative estimate of 1 day is adopted} \citep{2016PASP..128d4101Y}. The typical $\gamma_{\rm min}=40$ (e.g., \citealt{2014ApJS..215....5K}; \citealt{2014ApJ...788..104Z}) and $\gamma_{\rm max}=1\times10^{8}$ ($\gamma_{\rm max}>>100\gamma_{\rm b}$) are adopted in our fitting, which will not affect our main results (e.g., \citealt{2016MNRAS.461.1862K}). The other parameters, $B$, $\delta$, $p_{1}$, $p_{2}$, $\gamma_{\rm b}$, $N_{0}$ and $U_{\rm BLR}$, were kept free in our fitting.

 The multi-wavelength SEDs of GB 1310+487 are reproduced using the least-square ($\chi^{2}$) fitting technique (e.g. \citealt{2011ApJ...733...14M};
 \citealt{2012ApJ...752..157Z,2014ApJ...788..104Z}; \citealt{2014ApJS..215....5K}). In order {to constrain the} model parameters (e.g., $B$) by the synchrotron radiation spectrum, the GHz (86.24 and 142.33 GHz, see Table 5 in \citealt{2014A&A...565A..26S}) radio data are also included in $\chi^{2}$ fitting where a slow rising trend in the radio before and during the $\gamma$-ray flares occurred may suggest a common origin of the GHz radio and $\gamma$-ray emission, as suggested for other blazars (\citealt{2009ApJ...707L..56K}; \citealt{2011ApJ...741...30A}; \citealt{2012A&A...537A..32A}; \citealt{2012ApJ...744..177L}; \citealt{2012ApJ...758...72W}; \citealt{2013MNRAS.431.2481D}; \citealt{2013MNRAS.428.2418O}), so it may be reasonable that the radio data is included in the SED fitting. There are 25 observational data points (including 2 radio, 9 optical and infrared, 1 X-ray and 13 $\gamma$-ray data points) in Flare 1 state, 37 observational data points (including 2 radio, 24 optical and infrared, 1 X-ray and 10 $\gamma$-ray data points) in Flare 2 state and 11 observational data points (including 2 radio, 5 optical and infrared, 1 X-ray and 3 $\gamma$-ray data points) in Post-Flare state in the SED modeling. The observational error of the {data points} in the radio, infrared, optical, X-ray and $\gamma$-ray band are considered in $\chi^{2}$ fitting.
 We generate all the parameters in a broad range, and calculate the reduced $\chi^{2}_{\rm r}$ for these parameters. Then we derive the probability distribution of $\chi^{2}_{\rm r}$ (e.g., $p\propto\exp(-\chi_{\rm r}^{2})$), and the maximum probability {corresponds to the best-fit} parameters. The $1\sigma$ uncertainty of each parameter is derived from the Gaussian fit to its probability distribution by setting other {parameters to their best-fit values} (e.g., \citealt{2012ApJ...752..157Z,2014ApJ...788..104Z}; \citealt{2014ApJS..215....5K}).

The best fits are shown in Figure \ref{fig:flare}. The dotted, dashed, dot-dashed, long-dashed and solid lines represent the synchrotron, SSC, $\rm EC_{BLR}$, $\rm EC_{torus}$ and total emission respectively. The 1-sigma parameter spaces are shown with the gray background inside the plot. The upper, middle and lower panels show the SEDs of Flare 1, Flare 2 and Post-flare state respectively. 
{The higher $\chi^2$ value of Flare 1 indicates a worse fit compared to Flare 2 and Post-flare (see Table \ref{tab:GB1310p487}), due to the bad fit of the model for the high energy points.} The best-fit parameters, uncertainties and the values of $\chi^2$ are listed in Table \ref{tab:GB1310p487}. We find that the SEDs of Flare 1, Flare 2 and Post-flare state can be roughly reproduced by the leptonic jet model with the Syn + SSC + EC (BLR) + EC (torus) model.

From Flare 1, Flare 2 to the post-flare, the magnetic field intensities B are gradually increasing from 0.40$\pm$0.02, 0.70$\pm$0.04 to 0.76$\pm$0.05; $\gamma_{\rm b}$ decreases from $2.07\pm$0.07, 1.93$\pm$0.05 to 1.50$\pm$0.08; and  $N_{0}$ decreases from 1.11$\pm$0.32, 0.43$\pm$0.35 to 0.36$\pm$0.15, which are consistent with the the results obtained by \cite{2016PASP..128d4101Y}. However, the Doppler factor $\delta$ is a {rough constant which agrees well} with the Doppler factor estimated from the variability brightness temperature by assuming the intrinsic brightness temperature is limited to an equipartition value (e.g., see \citealt{{2009A&A...494..527H}} ; \citealt{2009PASJ...61..639F, 2009RAA.....9..751F}; \citealt{2010A&A...512A..24S} for more details). {This is inconsistent with} the results obtained by \cite{2016PASP..128d4101Y} {where it gradually decreases.} We also note an interesting result that the energy density of {the} BLR ($U_{\rm BLR}$) in our modeling is gradually {decreased} from From Flare 1, Flare 2 to the post-flare, where $U_{\rm BLR}=(1.55\pm0.47)\times10^{-3}~{\rm erg~cm}^{-3}$ (Flare 1), $U_{\rm BLR}=(1.48\pm0.26)\times10^{-3}~{\rm erg~cm}^{-3}$ (Flare 2), $U_{\rm BLR}=(0.21\pm0.06)\times10^{-3}~{\rm erg~cm}^{-3}$ (Post-flare) in rest frame.

\begin{table}
\centering
%\tiny
\caption{The relevant parameters of GB 1310+487 (input model parameters and output luminosities).}\label{tab:GB1310p487}
\begin{tabular}{lcccc}
\hline \hline
Parameter\tablenotemark{a}           &  Flare 1             &   Flare 2           &  Post-Flare     \\
\hline
$B$	                            &0.40$\pm$0.02         &0.70$\pm$0.04        &0.76$\pm$0.05   \\
$\delta$  	                    &15.97$\pm$0.46        &16.32$\pm$0.37       &16.17$\pm$0.55     \\
$p_{1}$  	                    &2.11$\pm$0.02         &2.15$\pm$0.04        &2.14$\pm$0.03       \\
$p_{2}$  	                    &14.02$\pm$3.23        &14.24$\pm$3.54       &14.21$\pm$4.46       \\
$\gamma_{\rm b}$    	        &$2.07\pm$0.07         &1.93$\pm$0.05        &1.50$\pm$0.08       \\
$N_{0}$                	        &1.11$\pm$0.32        &0.43$\pm$0.35       &0.36$\pm$0.15    \\
$U_{\rm BLR}$                   &1.55$\pm$0.47         &1.48$\pm$0.26        &0.21$\pm$0.06   \\
$\chi^2_{\rm r}$  	            &3.12                   &1.79                 &0.98                     \\
$L_{B}$                         &$2.63\times10^{44}$   &$8.07\times10^{44}$  &$9.51\times10^{44}$   \\
$L_{\rm e}$                     &$1.42\times10^{45}$   &$5.60\times10^{44}$  &$4.49\times10^{44}$   \\
$\epsilon_{B}$  	            &0.18                  &1.46                 &2.12                 \\
$r_{\rm diss}$                    &2.64$R_{\rm BLR}$    &2.68$R_{\rm BLR}$   &5.20$R_{\rm BLR}$  \\
\hline \hline
\end{tabular}
\tablenotetext{a}{{The dimensional of model parameters are:} \\
$B$(G), $\gamma_{\rm b} (10^{4})$, $N_{0} (10^4~{\rm cm}^{-3})$, $U_{\rm BLR} (10^{-3}{\rm erg~cm}^{-3})$,
$L_{B} (\rm erg~s^{-1})$, $L_{\rm e} (\rm erg~s^{-1})$, $\epsilon_{B}=L_{B}/L_{\rm e}$, }
\end{table}

\begin{figure}[h]
%\centering
\includegraphics[width=9cm,height=6.15cm]{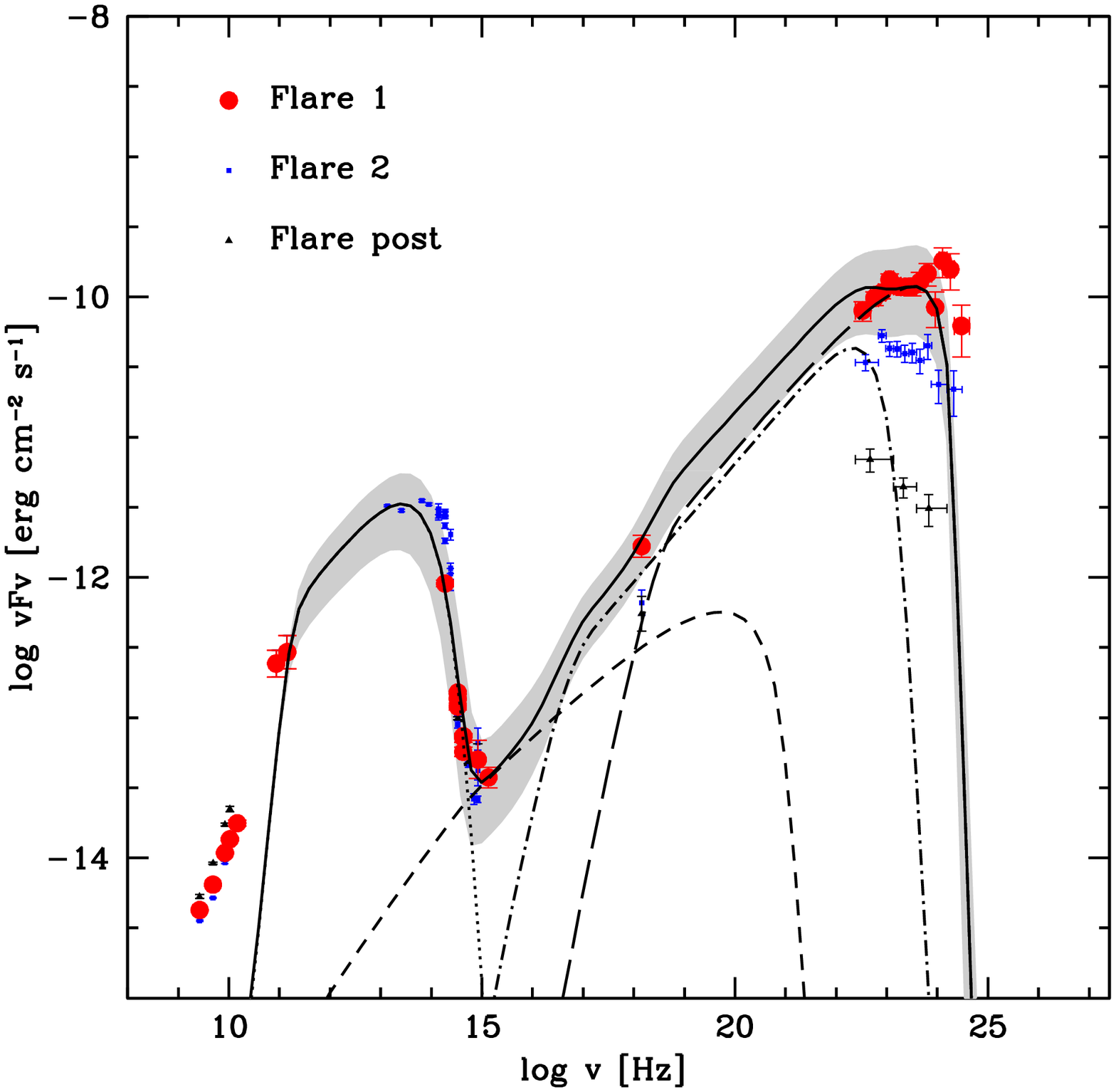}
\includegraphics[width=9cm,height=6.15cm]{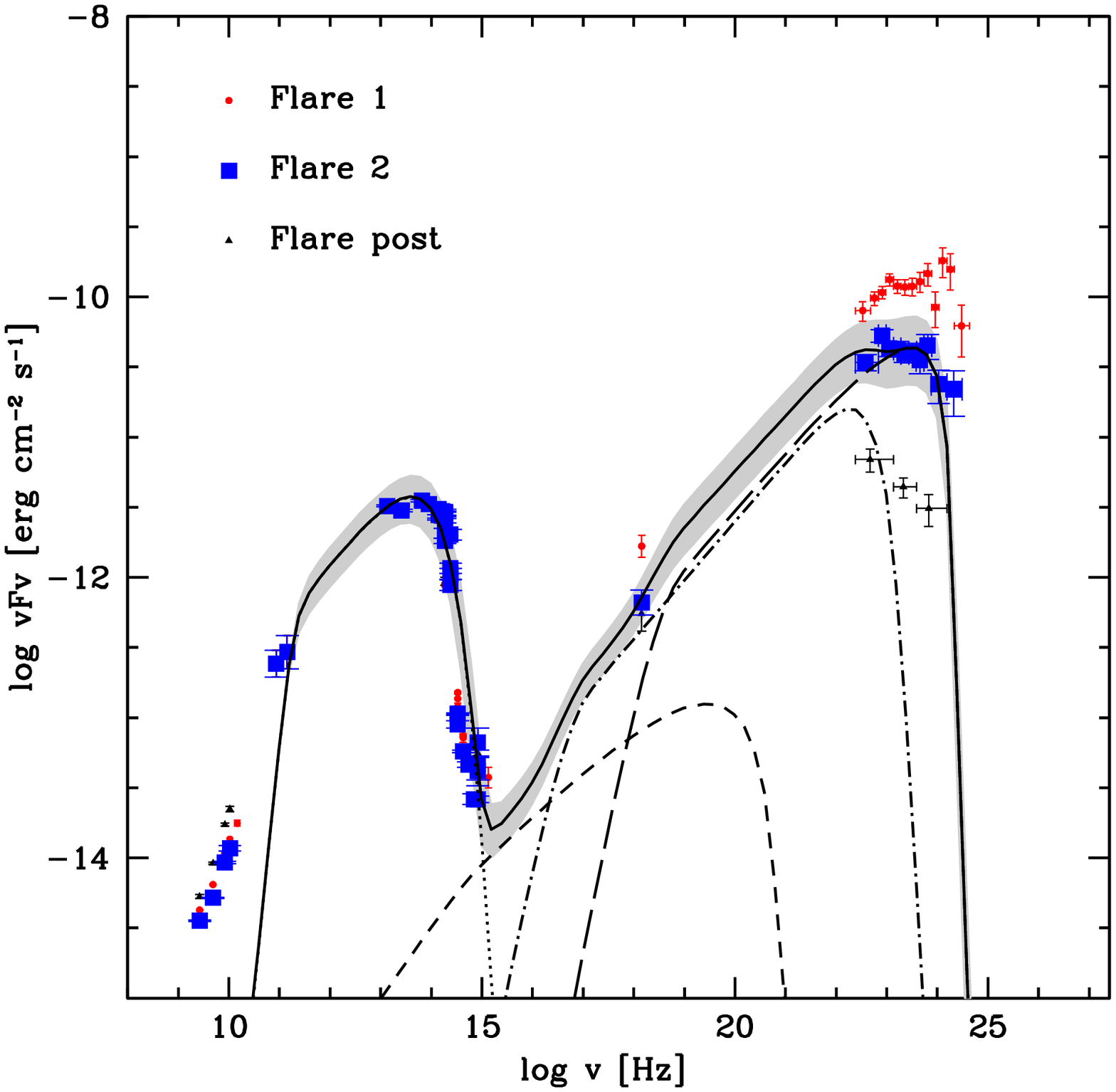}
\includegraphics[width=9cm,height=6.15cm]{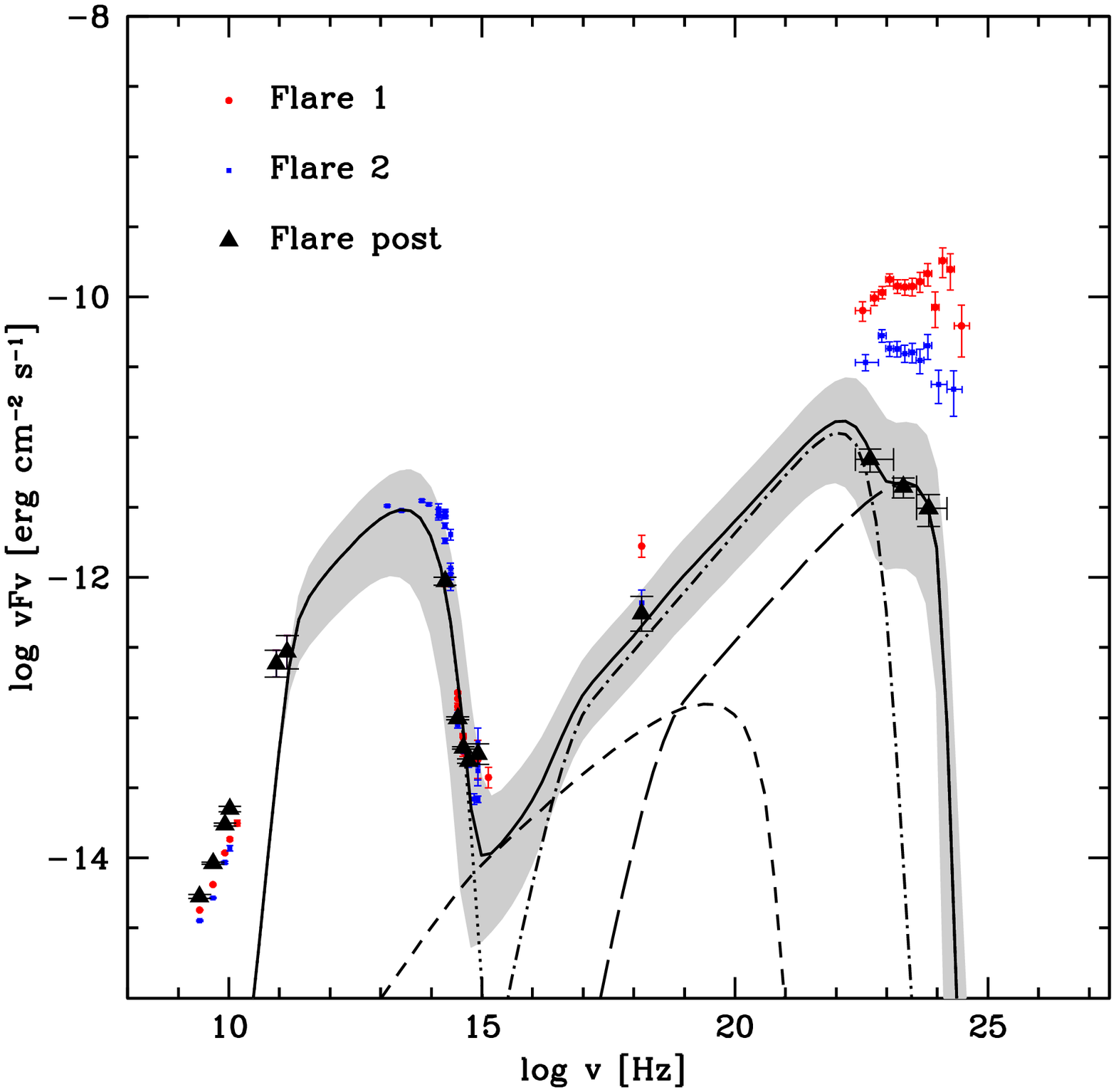}
\caption{The {SED} of GB 1310+487. The red solid points, blue squares and  black triangles indicate the broadband quit-simultaneous observational data of Flare 1, Flare 2 and Post-Flare state \citep{2014A&A...565A..26S}. The dotted, dashed, dot-dashed, long-dashed and solid lines represent the synchrotron, SSC, $\rm EC_{BLR}$, $\rm EC_{torus}$ and total emission. The gray backgrounds inside the plot indicate the 1-sigma parameter space of the SEDs model. The upper, middle and lower panels show the SEDs of Flare 1, Flare 2 and Post-Flare state respectively.}
\label{fig:flare}
\end{figure}

\section{Constrain $\gamma$-ray emission region site}\label{sec:region}

Recently, some works proposed that the energy densities of {the} BLR radiation ($U_{\rm BLR}$) are a function of dissipation distance $r_{\rm diss}$ from the central black hole (e.g., \citealt{1996MNRAS.280...67G}; \citealt{2009MNRAS.397..985G}), which can be approximately expressed {by} (see, \citealt{2009ApJ...704...38S}; \citealt{2012ApJ...754..114H})
\begin{equation}\label{eq3}
  U_{\rm BLR}(r)=\frac{\tau_{\rm BLR} L_{\rm disc}}{4\pi R^2_{\rm BLR}c[1+(r_{\rm diss}/R_{\rm BLR})^3]},\
\end{equation}
The reverberation mapping indicated that the typical size of {the} BLR is related to the disc luminosity $L_{\rm disc}$: $R_{\rm BLR}=10^{17}(L_{\rm disc}/10^{45}\rm \ erg\ s^{-1})^{1/2}\ $cm (e.g., see \citealt{2007ApJ...659..997K}; \citealt{2009ApJ...697..160B}; \citealt{2009MNRAS.397..985G}; \citealt{2014Natur.515..376G}). The equation (\ref{eq3}) can be {rewritten} as (see, \citealt{2015MNRAS.454.1310Y})
\begin{equation}\label{eq4}
U_{\rm BLR} (r)\simeq\frac{0.3\tau_{\rm BLR}}{1+(r_{\rm diss}/R_{\rm BLR})^3}\rm \ erg\ cm^{-3},\
\end{equation}
where $\tau_{\rm BLR}$ is the fraction of the disc luminosity reprocessed into BLR radiation. The typical value is $\tau_{\rm BLR}=0.1$ (e.g., see \citealt{2014Natur.515..376G}).

{Using equation} (\ref{eq4}), the distance $r_{\rm diss}$ from the central black hole to the emitting blob can be calculated as $r_{\rm Flare 1} \simeq 2.64 R_{\rm BLR}$, $r_{\rm Flare 2} \simeq 2.68 R_{\rm BLR}$ and $r_{\rm Post-Flare} \simeq 5.20 R_{\rm BLR}$ in Flare 1, Flare 2 and the Post-Flare states respectively, based on the $U_{\rm BLR}$ obtained from model fitting the SEDs of GB 1310+487. The average value of the distance is $r_{\rm average} \simeq 3.51 R_{\rm BLR}$. Which is consistent with some recent work in some blazars (e.g., \citealt{2010ApJ...714L.303F}; \citealt{2014ApJ...782...82D}; \citealt{2015MNRAS.454.1310Y}; \citealt{2015ApJ...803...15P}).

The luminosity of BLR ($L_{\rm BLR}\simeq 1.08\times10^{41}$ erg~$s^{-1}$) can be obtained from the luminosity ($L_{\rm H\beta}$) of the ${\rm H\beta}$ emission line of GB 1310+487 (Flux of ${\rm H\beta}$ line is $(0.24\pm0.06) \times 10^{-17}$ $\rm erg~cm^{-2} s^{-1}$, see  \citealt{2014A&A...565A..26S}), {base on equation 1} in \cite{1997MNRAS.286..415C}. Assuming the disc luminosity $L_{\rm disc} \simeq 10 L_{\rm BLR}$, we estimate the size of BLR ($R_{\rm BLR} \simeq 3.28\times10^{15}$ cm) and the distance from the central black hole to the emitting blob $r_{\rm diss} \simeq 3.7 \times 10^{-3}$ pc.

\section{Conclusion and Discussion}{\label{sec:DC}}

   In this work, we employ a leptonic model with the least-square ($\chi^{2}$) fitting technique to reproduce the multi-wavelength SEDs of GB 1310+487 in Flare 1, Flare 2 and  Post-Flare states.  The leptonic jet model with the Syn + SSC + EC (BLR) + EC (torus) model can reproduce the SEDs of GB 1310+487 in Flare 1, Flare 2 and  Post-Flare states, where the magnetic energy increase, matter energy decrease and external photon fields decrease. The dissipation distance $r_{\rm diss}$ from the $\gamma$-ray emitting region to the central black hole is constrained at $r_{\rm average} \simeq 3.51 R_{\rm BLR}$, from $r_{\rm Flare 1} \simeq 2.64 R_{\rm BLR}$ (Flare 1), $r_{\rm Flare 2} \simeq 2.68 R_{\rm BLR}$ (Flare 2) to $r_{\rm Post-Flare} \simeq 5.20 R_{\rm BLR}$ (Post-Flare). We propose that {the transformation of the magnetic field, the matter energy} and the location of the $\gamma$-ray emission region (or ambient photon field) may play an important role in different radiation states of GB 1310+487.

   From Flare 1, Flare 2 to Post-Flare, the ratio of magnetic energy and emitting-electron energy in the blob $\epsilon_{B}=L_{B}/L_{\rm e}$ (see Table \ref{tab:GB1310p487}) are gradually increasing from $\epsilon_{B}=0.18$, $\epsilon_{B}=1.46$ to $\epsilon_{B}=2.12$, which are consistent with other Blazars with near energy equipartition (e.g., \citealt{2014Natur.515..376G}). The magnetic energy increase and matter energy decrease, which suggest an effective acceleration of the emitting electrons takes place at the expense of energy of the magnetic field. It might indicate that the reduced energy of the Poynting flux is used to accelerate electrons (\citealt{2016PASP..128d4101Y}). Which {may} be one of the major factors to induce {the observed activity of the  GB 1310+487} by a conversion from magnetic energy to the energy of the {radiating electrons}.

  The input energy density of external photon fields $U_{\rm BLR}\sim(0.2-1.5)\times10^{-3}~{\rm erg~cm}^{-3}$ in flare 1, flare 2 and Post-Flare states are gradually decreasing and about 1-2 orders of magnitude lower than that in luminous FSRQs, where $U_{\rm BLR}\sim2.6\times10^{-2}~{\rm erg~cm}^{-3}$ (see \citealt{2008MNRAS.387.1669G}; \citealt{2009MNRAS.397..985G}, for details). This may be caused by the variation of the location of the $\gamma$-ray emission region, based on the energy densities of BLR radiation ($U_{\rm BLR}$) are a function of dissipation distance $r_{\rm diss}$ from the central black hole (e.g., \citealt{1996MNRAS.280...67G}; \citealt{2009MNRAS.397..985G}),
{or are caused} by a decreasing ambient photon field that might be caused by a decreasing accretion rate onto the central supermassive black hole \citep{{2011ApJ...736..128P}}. The increase of the flux of inverse Compton emission not accompanied by the increase of the flux of synchrotron emission as observed in Flare 1 and Flare 2 states may suggest this viewpoint. However, it should be noted that the high energy $\gamma$-ray spectrum in Flare 1 couldn't be well reproduced, which may be caused by different (or extra) radiation mechanism {between} Flare 1 and Flare 2. Flare 1 shows a fast rise and slower decay trending, while a gradual flux rise and rapid decay was observed in Flare 2 (see, \citealt{2014A&A...565A..26S}). For the various kinds of flux evolution of the two flares, further research is needed to {explore different generation mechanisms.}

  Based on the link of the energy densities of BLR radiation ($U_{\rm BLR}$) and the dissipation distance $r_{\rm diss}$ from the central black hole (e.g., \citealt{1996MNRAS.280...67G}; \citealt{2009MNRAS.397..985G}; \citealt{2009ApJ...704...38S}; \citealt{2012ApJ...754..114H}), we calculated the dissipation distance $r_{\rm Flare 1} \simeq 2.64 R_{\rm BLR}$ (Flare 1), $r_{\rm Flare 2} \simeq 2.68 R_{\rm BLR}$ (Flare 2) and $r_{\rm Post-Flare} \simeq 5.20 R_{\rm BLR}$ (Post-Flare), with a average value $r_{\rm average} \simeq 3.51 R_{\rm BLR}$. The value of the dissipation distance $r_{\rm diss}$ is roughly consistent with some recent works (e.g., \citealt{2010ApJ...714L.303F}; \citealt{2014ApJ...782...82D}; \citealt{2015MNRAS.454.1310Y}; \citealt{2015ApJ...803...15P}) that the $\gamma$-ray emission region of blazar jet might be located near the outer boundary of the BLR and within the dust tours (e.g., {\citealt{2012A&A...537A..70S,2013ApJ...773..147J}}; \citealt{2013ApJ...771L...4C}; \citealt{2014ApJ...782...82D}; {\citealt{2015ApJ...808..162C}}; \citealt{2016ApJ...821..102B};   \citealt{2017ApJS..228....1Z})
where contributions from both BLR and torus photons are required to explain the observed gamma-ray spectrum. 
{It should be noted that the $R_{BLR}$ of GB 1310+487 is much less than that of other typical blazars (e.g., $R_{BLR}\sim0.1$ pc), and, therefore, 
the $r_{\rm diss} \simeq 3.51 R_{BLR}$ is very small. It may be {that} the $R_{BLR}$ is underestimated due to the low flux of {the ${\rm H\beta}$ line} (e.g., contaminated by the foreground galaxy) with $(0.24\pm0.06) \times 10^{-17}$ $\rm erg~s^{-1}~cm^{-2}$ \citep{2014A&A...565A..26S}, or some other reasons, for instance, 
 it would be the case if the central black hole mass ($M_{\rm BH}$) is smaller than the one typically found in blazars, since the {$R_{BLR}$ gros with increasing} $M_{\rm BH}$ (e.g., \citealt{1999ASSL..234..157H,1999ApJ...526..579W}).
 In addition, one other thing to be noted is that  {GB 1310+487 is located in a} double system,  a foreground galaxy at z=0.500 (probably not AGN) and the background AGN at z=0.638 (Sokolovsky et al. 2014). The absorption of the foreground galaxy would result in the optical spectrum ``redder" (thus forming the steep spectrum) and the X-ray spectrum ``harder". Which complicates the interpretation of the SED, the model fitting parameters might be affected by the absorption of the foreground galaxy, especially in the optical part of the spectrum (e.g., the larger than usual value of $p_2$)(e.g., \citealt{2014A&A...565A..26S,2016PASP..128d4101Y}).
 The large number of free parameters may also affect the model {parameters constraints}, particularly in the {case of a possible} degeneracy in the model; or the model used in the work {is a too simplistic model for this problem.}
Our results suggest that the location of the $\gamma$-ray emitting region of GB 1310+487 is tightly constrained at the outer boundary of the BLR (the dissipation distance of the emission region from central black hole  $r_{\rm diss} \sim {\rm~a~few~times~of}~R_{\rm BLR}$), where both BLR and torus energy densities are contributed to the observed $\gamma$-ray spectrum. {The conversion of the magnetic field and the matter (radiation electrons) energy and the location of the $\gamma$-ray emission region (or ambient photon field) may play an important role in different radiation states of GB 1310+487.}

%%%%%%%%%%%%%%%%%%%%%%%%%%%%%%%%%%%%%%%%%%%%%%%%%%%%%%%%%%%%%%
%{\color{red}\rule{8cm}{1mm}}
\acknowledgments
We thank the anonymous referee for very constructive and helpful comments and suggestions, which greatly helped us to improve our paper. This work is supported by the Research Foundation for Advanced Talents of Liupanshui Normal University (LPSSYKYJJ201506). %This work is aslo supported by the Research Foundation of Liupanshui Normal University (LPSSYDXS1514, LPSSY201401), the Natural Science Foundation of the Department of Education of Guizhou Province (QJHKYZ[2015]455).

\end{document}